\documentstyle[preprint,tighten,aps,prl,epsfig]{revtex} 

\begin{document} 

\preprint{TRI-PP-97-31} 
\draft 
\title{Nucleon-nucleon bremsstrahlung: An example of the impossibility of
measuring off-shell amplitudes} 

\author{Harold W. Fearing}
\address{TRIUMF, 4004 Wesbrook Mall, Vancouver, B.C., Canada V6T 2A3}
\date{\today}
\maketitle
\begin{abstract}
For nearly fifty years theoretical and experimental efforts in nucleon-nucleon
bremsstrahlung (NN$\gamma$) have been devoted to measuring off-shell amplitudes
and distinguishing among various NN potentials on the basis
of their off-shell behavior. New experiments are underway, designed
specifically to attain kinematics further off shell than in the past, and thus
to be more sensitive to the off-shell behavior.  This letter shows that,
contrary to these expectations, and due to the invariance of the
S-matrix under transformations of the fields, the off-shell NN amplitude is
{\em as a matter of principle} an unmeasurable quantity in NN$\gamma$.

\end{abstract}
\narrowtext

The nucleon-nucleon bremsstrahlung (NN$\gamma$) reaction, specifically $p + p
\rightarrow p + p + \gamma$, was originally proposed \cite{Ashkin49} as a way
of determining off-shell aspects of the nucleon-nucleon (NN) force and
distinguishing among different NN potentials. Since the original suggestion,
there have been many calculations,
Refs.~\cite{Fearing87,Theory,Jetter95,Eden96}, mostly in non-relativistic
potential models. 
Recent experiments, e.~g.\cite{Michaelian90}, unlike earlier ones, seem to
show that within the context of contemporary potential models, off-shell
effects are important.  However most modern potentials seem to have similar
off-shell behavior in the region which has been explored, and thus it has not
yet been possible to distinguish among potentials.  

New experiments underway \cite{Newexpt97} will provide data in new
kinematic regions and with much better accuracy than before.
While there have been some new motivations to look at NN$\gamma$, notably the
fact that it may be a useful probe of heavy ion reactions,
these new experiments have all been designed to explore kinematics further off
shell than before so that they will be more sensitive to off-shell
effects. Thus a primary aim has been to distinguish among potentials via their
off-shell behavior, in accord with the established expectations for NN$\gamma$.

The aim of this paper is to show that, in contrast to these expectations,
in actual fact the off-shell NN amplitude is {\em unmeasurable} in
NN$\gamma$. This is true {\em in principle} and follows from the invariance of
the S-matrix under transformations of the fields. This result has profound
implications for it means that much previous work on NN$\gamma$, 
aimed at determining the off-shell NN amplitude and
distinguishing among potentials, was in fact misguided.
The principle is general, and thus NN$\gamma$ here serves as an example of the 
unmeasurability of off-shell effects.

To make a convincing case for this result it is necessary to understand
in a qualitative way the ambiguities of the usual non-relativistic potential
model approach.  One must also understand why this
result has not been noticed in the past.  It turns out that inherent
approximations in potential models camouflage the effect. Hence to proceed
further, a simple field theory model is described in which everything can be
calculated and in which one can see rigorously exactly how the off-shell NN
amplitude enters and why it is not a measurable quantity.

\begin{figure}[ht]
\caption{(a) External radiation graph off shell at the strong vertex;
(b) The double scattering contribution;
(c) A contribution off shell at the electromagnetic vertex;
(d) The contact term.
\label{fig:ext}}

\begin{center}
\begin{minipage}[ht]{85mm}
\epsfig{file=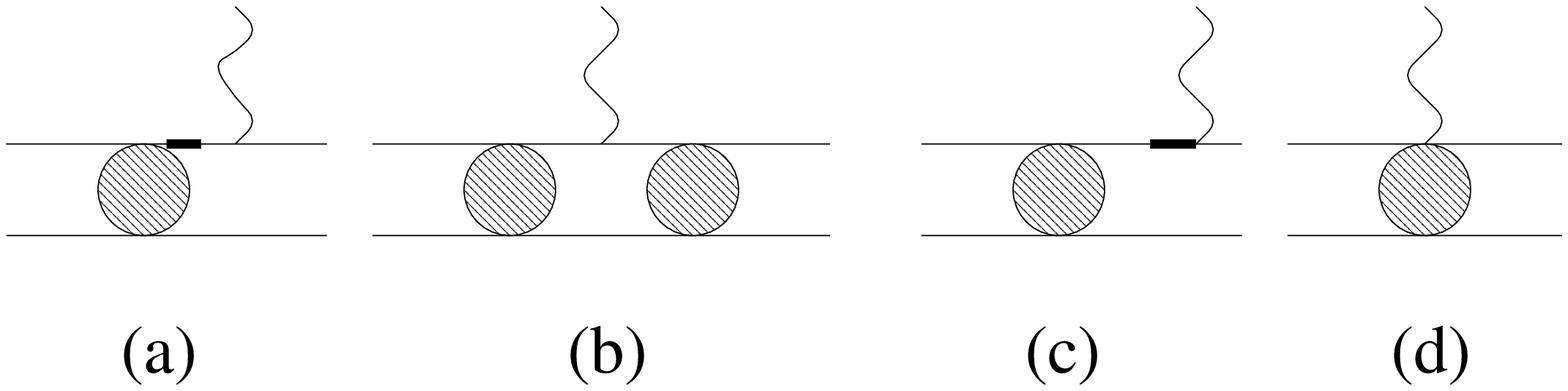,width=8cm}
\end{minipage}
\end{center}

\end{figure}

In the context of a non relativistic potential model one can rigorously define
and calculate, say by solving the Lippman-Schwinger equation, both an on- and
off-shell NN amplitude. To connect this
off-shell amplitude  with NN$\gamma$ experiments one uses it in the external
radiation graph in which a photon is simply attached to the external legs, as
in Fig.~\ref{fig:ext}a and its permutations.  Modern calculations also include
the so-called double scattering contribution, Fig.~\ref{fig:ext}b which is
typically a 10-30\% contribution.  The full NN$\gamma$ amplitude also involves a
piece with off-shell effects in the electromagnetic interaction, as in
Fig.~\ref{fig:ext}c, which has been considered by only a few authors,
\cite{Nyman}.

\begin{figure}[ht]
\caption{A typical contribution to the strong potential (a), leading to a
usually neglected contribution (b) to the contact term of the NN$\gamma$
amplitude. \label{fig:cont}}

\begin{center}
\begin{minipage}[ht]{85mm}
\epsfig{file=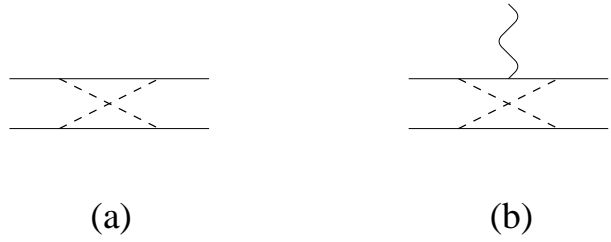,width=8cm}
\end{minipage}
\end{center}

\end{figure}

Another contribution is a contact term, Fig.~\ref{fig:ext}d, which reflects
radiation from the charged lines interior to the strong interaction and 
is necessary to preserve gauge invariance.  A simple example of such a
contribution is given in Fig.~\ref{fig:cont}. In a potential model it is
impossible to calculate such contributions rigorously. Even the
most microscopic potentials have a large number of phenomenological components,
e.~g. form factors, for which the underlying currents are not known.  One can 
fix part of the leading contribution to this
contact term by some variant of a soft photon approximation, and some explicit
exchange contributions can be calculated \cite{Jetter95,Eden96,Jong95}. However
in most potential model calculations most contributions of this contact
term are just neglected.

Before proceeding to more rigorous analysis it is important to get a
qualitative understanding of the reasons off-shell effects are
unmeasurable. The off-shell NN amplitude can be written schematically as
$T=T_{on} + (p^2-m^2) T_{off}$ where $T_{on}$ is the on-shell amplitude and
$(p^2-m^2) T_{off}$ is the off-shell part which vanishes when the particle goes
on shell, $p^2 = m^2$.  When this off-shell amplitude is put into the
bremsstrahlung amplitude of Fig.~\ref{fig:ext}a however, the $(p^2-m^2)$ factor
is exactly cancelled by the propagator. Thus the part 
involving $T_{off}$ looks just like a contact term and could have been included
in the contact term. So there is an obvious ambiguity in that off-shell effects
could be considered part of the external radiation diagram or part of the
contact term.

A second important observation arises from the field theory result
\cite{Kamefuchi61} that it is possible to make transformations on the fields in
the theory, which will change the interaction, but will not change any physical
quantity, in particular any S-matrix element.

The third observation is that it is the terms in the interaction proportional
to the equation of motion (EOM) which are changed by such
transformations \cite{Weinberg}.  In the present case the EOM generates (in
momentum space and for spin zero particles) the off-shell factor
$(p^2-m^2)$. Thus field transformations in effect can arbitrarily change the
coefficient of the off-shell contribution without changing any of the physical
results.

Now it is possible to understand in a qualitative sense the problem
with the off-shell contributions. At the simplest level there will always be an
ambiguity in these off-shell terms because they can also be written as contact
terms. Usually such contact terms are dropped, so one gets
different off-shell dependences depending on how much of the off-shell term is
included in the terms kept.

At a more fundamental level one can always make a transformation of the fields
which changes the coefficient of the EOM terms in the interaction and hence the
coefficient of the off-shell NN contributions to the NN$\gamma$ amplitude
(Fig.~\ref{fig:ext}a). Since this transformation does not change the S-matrix
it does not change the measurable NN$\gamma$ amplitude. Thus a measured
NN$\gamma$ amplitude corresponds to any of an infinite selection of
coefficients of the off-shell part of the amplitude. It thus follows that such
off-shell amplitudes are simply not measurable.

Furthermore since the NN$\gamma$ amplitude cannot depend on the transformation,
it cannot depend at all on the coefficients of the EOM terms in the Lagrangian, 
since one
possible transformation would make all such coefficients zero. Thus such terms
must cancel exactly between off-shell parts appearing in the diagrams of
Fig.~\ref{fig:ext}a and the contact term of Fig.~\ref{fig:ext}d.

There may also be off-shell contributions from parts of the interaction which
do not come directly from the EOM terms in the Lagrangian. Such terms 
cannot be separated in the
amplitude from the EOM terms and so cannot be separately measured. Furthermore
they come from pieces of the interaction which are already determined by
on-shell information only.

It is interesting to speculate on why this scenario has not been realized in
the past.  In part it must be because it depends crucially on an understanding
of cancellations between the main terms and the contact term, which is essentially 
impossible to calculate in potential models and so is usually dropped at
the very beginning. Furthermore, the concept of field transformations and
invariance of the S-matrix under such transformations is really a field theory
concept somewhat distant from the concepts familiar in non relativistic
potential model calculations.  There have been however a few similar
observations for related processes. For example Scherer and Fearing
\cite{Scherer95a,Scherer95b} showed that off-shell effects did not contribute
to Compton scattering on the pion and others \cite{Davidson96} have discussed
similar concepts with respect to the pion-nucleon system.

For completeness note that off-shell vertices are often simply parameterized,
rather than calculated from the interaction. These parameters become a
prescription accounting for unknown contact terms and other
approximations in the model, and clearly can be obtained from measurable
results. However they bear no direct relation to the 'physical' off-shell
amplitudes calculated directly from the interaction which are being discussed
here. 
 
To make this qualitative understanding more rigorous, consider a model which
allows a complete calculation of the contact terms and in which it is possible
to look at the effects of field transformations explicitly. The standard
potential model approach allows neither. The model is a simple field theory
for spin zero particles, only one charged. With one minor exception to
be discussed below, spin is really irrelevant. Thus the prototype reaction is
$\pi^+ + \pi^0 \rightarrow \pi^+ + \pi^0 + \gamma$. Take as the Lagrangian for
this reaction ${\cal L}={\cal L}_2+{\cal L}_4^{GL}+\Delta{\cal L}_4$, where
\begin{displaymath}
{\cal L}_2 = \frac{F_0^2}{4} \mbox{Tr} \left ( D_{\mu} U (D^{\mu}U)^{\dagger} \right)
+\frac{F_0^2}{4} \mbox{Tr} \left ( \chi U^{\dagger}+ U \chi^{\dagger} \right ),
\end{displaymath}
$D_{\mu}$ is a covariant derivative, $\chi$ involves the masses, $U$
contains the pion fields via $U =  {\exp}[i\tau \cdot \pi/{F_0}]$,
and ${\cal L}_4^{GL}$ is the usual Gasser-Leutwyler Lagrangian 
\cite{Gasser84}.  
$\Delta{\cal L}_4$ contains the two allowed EOM terms as 
described in Ref.~\cite{Rudy94}, namely
\begin{displaymath}
\Delta{\cal L}_4= 
\beta_1 \mbox{Tr}\left({\cal O}{\cal O}^{\dagger}\right) 
+\beta_2 \mbox{Tr}\left[(\chi U^{\dagger}-U\chi^{\dagger}){\cal O}\right].
\end{displaymath}
Here the ${\cal L}_2$ EOM is ${\cal O}=0 =
(D_{\mu}D^{\mu}U) U^{\dagger}-U(D_{\mu}D^{\mu}U)^{\dagger}
-\chi U^{\dagger}+U\chi^{\dagger}
+\frac{1}{2} \mbox{Tr}\left(\chi U^{\dagger}-U\chi^{\dagger}\right)$.

This is just the chiral perturbation theory Lagrangian in the meson sector
through $O(p^4)$. It should be strongly emphasized however that the results
here have nothing to do with chiral perturbation theory as such. ${\cal L}$
is just an effective Lagrangian which, through $O(p^4)$, 
is the most general one satisfying gauge invariance and
other appropriate symmetries. It is renormalizable to this order, but depends
on the parameters in ${\cal L}_4^{GL}$ and on $\beta_1$ and $\beta_2$.

Using this Lagrangian the fully renormalized, irreducible, amplitude for the
elastic process $\pi^+(p_1) + \pi^0(p_2) \rightarrow \pi^+(p_3) + \pi^0(p_4)$,
with the charged pions off shell is
\begin{eqnarray}
\label{eq:pipiamp}
\Gamma_{4\pi} &=& \frac{i}{F_0^2}[ T_0(p_1,p_3) - 
\frac{1}{3}(\Lambda_1+\Lambda_3)] +\frac{8 i m_\pi^2 \beta_2}{3 F_0^4} 
(\Lambda_1+\Lambda_3) \nonumber \\
&+& \frac{16 i \beta_1}{F_0^4}[T_0(p_1,p_3) (\Lambda_1+\Lambda_3) - \frac{1}{3}
(\Lambda_1^2+\Lambda_3^2)] ~~,
\end{eqnarray}
where
$T_0(p_1,p_3) \equiv T_0(p_1,p_2,p_3,p_4) = \frac{1}{2}[(p_1-p_3)^2 + (p_2-p_4)^2] 
- m_\pi^2$ with $\Lambda_i=p_i^2-m_\pi^2$. The tree level terms coming from 
${\cal L}_4^{GL}$ and the one loop terms involving ${\cal L}_2$ twice will 
never be written explicitly. Such terms are finite but irrelevant to 
the argument.

This amplitude involves an off-shell component, proportional to $\beta_1$ and
$\beta_2$, coming from the EOM terms. There are also a few off-shell
contributions from the first term, both proportional to $(\Lambda_1+\Lambda_3)$
and originating in the momentum dependence buried in $T_0$. However these terms
depend only on quantities available from the on-shell amplitude.

Now consider the most general field transformation to $O(p^2)$
\cite{Scherer95b} , $U \rightarrow {\exp}(iS) U$ where for arbitrary
real $\alpha_1, \alpha_2$
\begin{displaymath}
S=\frac{4i}{F_0^2}\left[\alpha_1{\cal O}
-\alpha_2\left(\chi U^{\dagger}-U \chi^{\dagger}-
\frac{1}{2}
\mbox{Tr}(\chi U^{\dagger}-U\chi^{\dagger})\right)\right].
\end{displaymath}
When applied to the fields $U$ in ${\cal L}_2$ it generates a correction to the
Lagrangian of ${\cal O}(p^4)$ given by $\delta{\cal L}_2 =\alpha_1 
\mbox{Tr}({\cal O}{\cal O}^{\dagger}) +\alpha_2 \mbox{Tr}
((\chi U^\dagger-U\chi^\dagger){\cal O})$,
which has exactly the same form as $\Delta{\cal L}_4$. Thus field
transformations can be used to change the coefficients $\beta_1$ and $\beta_2$
appearing in the original Lagrangian to zero or to
anything one wishes. Since such transformations do not effect the physical
results, the bremsstrahlung amplitude calculated with this Lagrangian, cannot
depend on the values of $\beta_1$ and $\beta_2$, which according to
Eq.~(\ref{eq:pipiamp}) are coefficients of off-shell parts of the
elastic amplitude.

To see this in more detail, use this Lagrangian to calculate the full
NN$\gamma$ amplitude.  The fully renormalized, irreducible electromagnetic
vertex for $\pi^+(p_i) \rightarrow \pi^+(p_f) + \gamma(k)$ is given by
\cite{Rudy94}
\begin{displaymath}
\Gamma_{\pi \pi \gamma} = -ie \epsilon \cdot (p_i+p_f) [ 1 +
\frac{16 \beta_1}{F_0^2}(\Lambda_i+\Lambda_f) ],
\end{displaymath}
where as before $\Lambda=p^2-m_\pi^2$.  Note the off-shell part $\sim \beta_1$.
There are also corrections to the renormalized pion propagator coming from
$\Delta{\cal L}_4$, \cite{Scherer95a}
\begin{displaymath}
\Delta^{-1}_R(p) = {(p^2 - m_\pi^2) [1+\frac{16 \beta_1}{F_0^2}(p^2-m_\pi^2)] 
+ i \epsilon }~.
\end{displaymath}

The corrections to the renormalized propagator cancel
exactly the off-shell part of the electromagnetic vertex. This is a general
result for spin zero particles and follows from the Ward-Takahashi identity
\cite{Ward50} which implies that at $k^2=0$ the the full renormalized propagator 
times the
half off-shell electromagnetic vertex is equal to the free propagator times the
on-shell electromagnetic vertex (c.f.~Appendix A of Ref.~ \cite{Rudy94}). Thus
for spin zero bremsstrahlung one never needs to consider off-shell effects at the
electromagnetic vertex.  For spin one half particles there may be terms in the
vertex involving magnetic moments which are gauge invariant by themselves and
so are not constrained by the Ward-Takahashi identity \cite{Ward50}. These may
give residual off-shell terms at the electromagnetic vertex which would be
treated exactly as those at the strong vertex.

Thus the amplitude for radiation from external legs is
\begin{eqnarray}
\label{eq:m3}
M_{1+3}&=& \frac{i e}{F_0^2} [T_0(p_1,p_3)-k \cdot (p_1-p_3)]
(\frac{\epsilon \cdot p_3}{k \cdot p_3}-\frac{\epsilon \cdot p_1}{k \cdot p_1})
\nonumber  \\ &-& \frac{2 i e}{3 F_0^2}  \epsilon \cdot (p_3+p_1) 
\nonumber   \\ & \times & \left[1 - \frac{48 \beta_1}{F_0^2}
[T_0(p_1,p_3)-k \cdot (p_1-p_3)]
- \frac{8 m_\pi^2 \beta_2}{F_0^2}\right] \nonumber  \\
&-& \frac{64 i e \beta_1}{3 F_0^4}(\epsilon \cdot p_3 k 
\cdot p_3 - \epsilon \cdot p_1 k \cdot p_1)  
\end{eqnarray}
The first term comes from the on-shell part of the strong amplitude.  The rest
arises from the off-shell part of the strong vertex, which cancels the
propagator denominator and thus looks like a contact term.

In this model the double scattering term, Fig.~\ref{fig:ext}b, would be a 
one-loop contribution $O(p^6)$.  Thus $M_{1+3}$ is the exact analogue of what is
calculated in a non relativistic potential model. It contains off-shell terms
depending on the parameters of the interaction $\beta_1$ and $\beta_2$ .  The
standard argument of the usual approach would be that by measuring the
NN$\gamma$ amplitude and comparing it with the calculated $M_{1+3}$ one could
determine $\beta_1$ and $\beta_2$ and thus distinguish among different
interactions.

This cannot be correct however since the field
transformations, which leave the measurable bremsstrahlung amplitude unchanged,
can change the coefficients $\beta_1$ and $\beta_2$ arbitrarily. Thus the
off-shell component of $M_{1+3}$, which in the spin one half case might come
also from the electromagnetic vertex, cannot be a measurable quantity.

The resolution of this problem depends on the renormalized contact term which can be
calculated explicitly in this model, unlike in the standard potential model
calculations. It is given by
\begin{eqnarray}
\label{eq:contact}
\Gamma_{4\pi\gamma}&=& \frac{2 i e}{3 F_0^2} \epsilon \cdot (p_1+p_3)  
\nonumber \\ & \times &  \left[ 1 -
\frac{48 \beta_1}{F_0^2}[T_0(p_1,p_3)-k \cdot (p_1-p_3)] 
- \frac{8 m_\pi^2 \beta_2}{F_0^2} \right] \nonumber  \\
&+&\frac{64 i e \beta_1}{3 F_0^4}(
\epsilon \cdot p_3 k \cdot p_3 - \epsilon \cdot p_1 k \cdot p_1)~~.
\end{eqnarray}

The $\beta_1$ and $\beta_2$ terms here cancel those in $M_{1+3}$
of Eq.~(\ref{eq:m3}) and the full amplitude is in fact independent of $\beta_1$
and $\beta_2$ as it must be. In this case even the single off-shell term
appearing in $\Gamma_{4\pi}$ which is independent of $\beta_1$ and $\beta_2$
cancels, though that is probably not a general result. The final amplitude thus
involves off-shell effects only via the momentum dependence implicit in $T_0$,
which is completely determined by the on-shell amplitude.

Some further insight can be obtained from the soft photon 
limit of the model amplitude, since it reflects
a model-independent imposition of gauge invariance, which enforces some of the
cancellation between external radiation and contact terms.  One sees from
Eq.~(\ref{eq:m3}) that the $O(1/k)$ part of $M_{1+3}$ is gauge invariant.
The rest of the first term, which has all the $O(k/k)$ parts, is also gauge 
invariant. This implies that the remaining
$O(k^0)$, independent of $k$ terms which are fixed by gauge invariance,
must be zero. That then enforces the cancellation of most of the off-shell
terms between this amplitude and the contact amplitude of
Eq.~(\ref{eq:contact}) leaving just a few $O(k)$ terms not fixed by gauge
invariance.

The results of this investigation can be summarized as follows. Field
transformations change coefficients of off-shell parts of the elastic
amplitude without changing the bremsstrahlung amplitude. Thus these off-shell
amplitudes are unmeasurable in NN$\gamma$, contrary to widespread expectations.
Invariance under these transformations means also that there will be large
cancellations between usual external radiation diagrams and the contact term,
which is normally not calculated in standard potential
approaches. Thus microscopic type calculations which allow a complete
evaluation of the contact term should be favored. At the very least, in
potential model calculations some technique which enforces gauge invariance and
the cancellations that entails, at least at soft photon order, should be
used. None of this means that NN$\gamma$ is less interesting. Quite the
contrary, since it is now clear that a full understanding of the process
requires a detailed understanding of the contact terms, which involve a photon
probe of the inner details of the NN interaction. Finally, 
none of these considerations are particularly specific to NN$\gamma$. It
is probable that off-shell amplitudes are unmeasurable in any process and
calculations purporting to show sensitivity of physically measurable quantities
to such off-shell effects should be viewed with suspicion.
 
The author would like to thank Stefan Scherer for a very careful reading 
of the manuscript, for some useful comments, and for what has now been some 
years of fruitful collaboration on these and related topics.

\end{document}